\documentclass[a4paper]{article}
\usepackage{graphicx, enumerate, subfig}
\usepackage{hyperref}
\hypersetup{colorlinks=true, citecolor=black, linkcolor=black, urlcolor=black, pdfpagemode=UseNone}
\usepackage{url}
\usepackage{amssymb, amsmath}
\usepackage{placeins} 
\usepackage{multirow}
\usepackage{geometry} 
\usepackage[rgb,table]{xcolor}
\usepackage{cleveref}
\usepackage[inline,shortlabels]{enumitem}
\usepackage[labelfont=bf]{caption}

\begin{document}

\begin{flushleft}
{\Large
\textbf{Nonlinear trend removal should be carefully performed in heart rate variability analysis}
}
\\
Binbin Xu$^{1,2}$, R\'{e}mi Dubois$^{1}$, Oriol Pont$^{2}$, Hussein Yahia$^{2}$
\\
\bf{1} IHU Institut de Rythmologie et Mod\'{e}lisation Cardiaque, Fondation Bordeaux Universit\'{e}, Bordeaux
\\
\bf{2} INRIA Bordeaux Sud-Ouest, Talence, France
\end{flushleft}

\subsubsection*{Abstract}
\begin{itemize}
\item \underline{Background} : In Heart rate variability analysis, the rate-rate time series suffer often from aperiodic non-stationarity, presence of ectopic beats etc. It would be hard to extract helpful information from the original signals.

\item \underline{Problem} : Trend removal methods are commonly practiced to reduce the influence of the low frequency and aperiodic non-stationary in RR data. This can unfortunately affect the signal and make the analysis on detrended data less appropriate.

\item \underline{Objective} : Investigate the detrending effect (linear \& nonlinear) in temporal / nonliear analysis of heart rate variability of long-term RR data (in normal sinus rhythm, atrial fibrillation, congestive heart failure and ventricular premature arrhythmia conditions).

\item \underline{Methods} : Temporal method : standard measure SDNN; Nonlinear methods : multi-scale Fractal Dimension (FD), Detrended Fluctuation Analysis (DFA) \& Sample Entropy (SampEn) analysis.

\item \underline{Results} : The linear detrending affects little the global characteristics of the RR data, either in temporal analysis or in nonlinear complexity analysis. After linear detrending, the SDNNs are just slightly shifted and all distributions are well preserved. The cross-scale complexity remained almost the same as the ones for original RR data or correlated. 

Nonlinear detrending changed not only the SDNNs distribution, but also the order among different types of RR data. After this processing, the SDNN became indistinguishable between SDNN for normal sinus rhythm and ventricular premature beats. Different RR data has different complexity signature. Nonlinear detrending made the all RR data to be similar, in terms of complexity. It is thus impossible to distinguish them. The FD showed that nonlinearly detrended RR data has a dimension close to 2, the exponent from DFA is close to zero and SampEn is larger than 1.5 -- these complexity values are very close to those for random signal.

\item \underline{Conclusions} : Pre-processing by linear detrending can be performed on RR data, which has little influence on the corresponding analysis. Nonlinear detrending could be harmful and it is not advisable to use this type of pre-processing. Exceptions do exist, but only combined with other appropriate techniques to avoid complete change of the signal's intrinsic dynamics.

\end{itemize}

\emph{Keywords}
\begin{itemize*}
\item heart rate variability 
\item linear / nonlinear detrending 
\item complexity analysis 
\item multiscale analysis 
\item detrended fluctuation analysis
\item fractal dimension
\item sample entropy;
\end{itemize*}

\section{Introduction}\label{sec_intro}%

Heart rate variability (HRV) has been since long time a standard method to evaluate the heart's performance. In 1996, the Task Force of the European Society of Cardiology and the North American Society of Pacing and Electrophysiology standardized the HRV measurement, its physiological interpretation and clinical use \cite{TaskForce1996Heart}. The fluctuation of intervals between normal heartbeats are mediated by autonomic inputs to the sinus node \cite{Stein1999Insights}. This means that by analyzing these fluctuations, information about the cardiac autonomic modulation and its changes can be obtained. HRV reflects these fluctuations, more precisely, the phasic modulation of heart rate. In pathological cases, the heart loses (part of) its central modulation capability of heart rate or there would be a lack of response of the sinus node. So, the HRV values will be lower than in normal case. 

There exists several ways to analyze the variability of heart rate. The basic ones are the traditional statistics in Time domain, like SDNN (the standard deviation of all N--N intervals; SDANN (the standard deviation of the average of N--N intervals). These measures have been clinically proven useful. Analysis can also be performed in Frequency-domain: high--frequency power; low--frequency power; very--low--frequency band; ultra--low--frequency band; total power \cite{Stein1999Insights,Voss2009Methods,RajendraAcharya2006Heart,Billman2011Heart}. Due to the strong nonlinear properties of this type of signals, methods of the third class based on nonlinear dynamics have been shown to be more robust \cite{Perkioemaeki2005Fractal}. They can be divided into several families \cite{Maestri2007Nonlinear,Ernst2014Methodological} : symbolic dynamics \cite{Guzzetti2005Symbolic,Cysarz2011Multiscale}, entropy \cite{Pincus1991Approximate,Richman2000Physiological}, fractality-multifractality \cite{Ivanov1999Multifractality,Gieraltowski2012Multiscale}, predictability \cite{Porta2007Complexity}, empirical mode decomposition \cite{Balocchi2004Deriving,Yeh2010Intrinsic}, and Poincar\'{e} plots \cite{Karmakar2011Sensitivity}. 

The physiological signals are often noised by perturbations which could arise from electrode changes due to perspiration, movement and respiration, from the electronic data acquisition systems themselves or be interferences from other organs. It is thus necessary to reduce these noises, especially in case of electrocardiological signals analysis where the baseline interferences are an unavoidable preprocessing step. Many techniques exist to remove these trends \cite{Cleveland1979Robust,Laguna1992Adaptive,Soernmo1993Time,Luo2010Review}. In HRV analysis, the RR time series suffer often from aperiodic non-stationarity, presence of ectopic beats. It would be hard to extract helpful information from the original signals. It is also necessary to perform the above-mentioned preprocessing -- trend removal which is commonly practiced \cite{Shin1989Assessment,Porges1990Analysis,Niemelae1994Effect,Litvack1995Time,Berntson1997Heart,Tarvainen2002Advanced,Thuraisingham2006Preprocessing}. 

The typical detrending in HRV analysis can be divided into two classes : linear (first order or higher polynomial model \cite{Litvack1995Time,Mitov1998Method}, moving polynomial model \cite{Porges1990Analysis,Berntson1997Heart}) and nonlinear detrending (smoothness priors method \cite{Tarvainen2002Advanced}, wavelet \cite{Thuraisingham2006Preprocessing}, wavelet packets \cite{Shafqat2007Evaluation}, nonparametric regression \cite{Cleveland1979Robust,Shin1989Assessment}). The objective remained the same : removing the low frequency and aperiodic non-stationary components and rejecting low periodic non-sinusoidal activity which may have higher frequency harmonics \cite{Porges1990Analysis}. However, this pre-processing step by detrending RR time series can affect the following analysis. In many cases, the influence can be dramatic. Comparison studies of the detrending influence in HRV analysis can be found in literature \cite{Yoo2004Effects,Thuraisingham2006Preprocessing,Shafqat2007Evaluation,Colak2009Preprocessing,Kaur2014Comparison}.The results are all interesting, but the conclusions often varied from on study to another or are sometimes contradictory. For example, detrending (linear or nonlinear) is suggested in study but not in another one; nonlinear analysis of detrended RR time series is not recommended in some studies, but is considered as a good practice to discriminate the different heart rate data in others'. Though in reality, all these results are true in their own dataset conditions (dependence of data samples, pathology) / methodology (methods choices, combination of other pre-processing steps), it is still a little confusing. 

From a point-view of pure signal processing, the linear detrending is tolerable in HRV analysis. Since linear detrending would not technically change too much intrinsic dynamics of the signal (especially in case of electrocardiological signals), it should not affect largely the analysis either in linear or nonlinear studies. But for nonlinear detrending, it should be performed carefully in both linear/nonlinear studies. In biomedical signal, it is the rich fluctuations which consists the true value of the signal. The nonlinear trend removal is in fact the same process of signal denoising. Removing the nonlinear trend, is somehow to remove the fundamental dynamics from the signal, what's left can be considered as ``noise''. Analyzing these ``nonlinearly detrended signal'' could lead to some wrong conclusions. If in some cases where the nonlinear detrending is required, the detrending should be appropriately determined. 

So, the objective of this work is to study the linear / nonlinear detrending effects on linear / nonlinear HRV analysis, with long-term \& large datasets for different pathologies to quantify the effects and to explain qualitatively the effects. 

\section{Data and Methods}\label{sec_method}

\subsection{Data}

The data used in HRV analysis can be on either short-term (like 500 seconds, 10 minutes, 1024 R-R intervals) or long-term data (24 hours). The former is most used because it is affordable in terms of time-cost and its results are relatively reasonable. Though it is supposed that the data acquisition would be performed in standard conditions, this is not often respected.  What's more, since the human heart rate / ECG could be influenced by many factors, such as emotion, exercise, daylight, sleep $\ldots$. the robustness became thus the main concern for short-term data analysis. Considering also the circadian rhythms, the long-term measure can provide more reliable data. So in this study, we used long-term data analysis ($1 \sim 1.2 \times 10^5$ samples). 

Four type of RR data are analyzed : normal sinus rhythm as baseline; three cardiac diseases : congestive heart failure, atrial fibrillation and ventricular premature arrhythmia. 
\begin{itemize}
\item Normal Sinus Rhythm (NSR) \cite{Stein1999Effect} : beat-to-beat data for 54 long-term (24 hours) ECG recordings of subjects in normal sinus rhythm. The original ECG recordings were digitized at 128 samples per second. The beat annotations were obtained by automated analysis with manual review and correction. \path{http://physionet.org/physiobank/database/nsr2db/}.

\item Atrial Fibrillation (AF) \cite{Petrutiu2007Abrupt} : The Long-Term AF Database includes 84 long-term ECG recordings of subjects with paroxysmal or sustained atrial fibrillation. Same data samples per second as for NSR and CHF database. \path{http://physionet.org/physiobank/database/ltafdb/}.

\item Congestive Heart Failure (CHF) \cite{Goldsmith1997Long} : beat-to-beat data for 29 long-term ECG recordings of subjects with congestive heart failure (NYHA classes I, II, and III). Digitized at 128 samples per second, these data have been manually reviewed and corrected. \path{http://physionet.org/physiobank/database/chf2db/}.

\item Ventricular Premature Arrhythmia (VPA) : from PhysioNet Cardiac Arrhythmia Suppression Trial (CAST) database comprising long-term ECG recordings of more than 800 patients \cite{Stein2000Clinical} (some data is missed, around 600 data are used). \path{http://physionet.org/physiobank/database/crisdb/}.
\end{itemize}

\subsection{Detrending Methods}

As above-mentioned, there exists many detrending methods, but the typical methods are linear and nonlinear detrending. For each category, the outputs will change little -- never dramatically. This is also the basic principle of every method, violating this rule will make the detrending be failed. So, in this work, we used to typical linear and nonlinear detrending methods :
\begin{itemize}
\item linear detrending : by computing the least-squares fit of a straight line to the data and subtracts the resulting function from the data which has been implemented in Matlab\textsuperscript{\textregistered} as a standard function. 
\item nonlinear detrending : the trend is often taken as the reconstructed signal from the wavelet decomposition \cite{Wiklund1997Short,Thuraisingham2006Preprocessing}. For comparative reason, we use the same method as in \cite{Thuraisingham2006Preprocessing} : the trend is the reconstruction of the sixth level from wavelet \texttt{db3} decomposition. 
\end{itemize}

\subsection{Analysis Methods}

In single scale analysis (directly on the original RR data), the standard HRV measure -- SDNN is used. For nonlinear analysis, we are interested in complexity analysis methods. It's known that special patterns or shifts can be often found in electrophysiological signals reflecting the system's dynamics. So, analyzing the complexity of these patterns can help to explore the searched physiological mechanisms. Three typical complexity analysis methods are used here : Fractal Dimension (FD), Detrended Fluctuation Analysis (DFA) and Sample Entropy (SampEn). We give only the principles of each method, the detailed algorithms can be found in original works. 

\begin{itemize}

\item Fractal Dimension. By quantifying their graph complexity as a ratio of the change in detail to the change in scale, Fractal dimension helps to measure the roughness or smoothness for time series or spatial data. Many FD estimators have been proposed : from the basic box-counting to variogram or by spectrum.\cite{Gneiting2012Estimators, Lopes2009Fractal}. But the principle remained the same: 
\begin{enumerate*}[(i)]
\item measure the quantities of the object using various step sizes; 

\item use least-squares regression to fit the graph (generally the log-log plot, measures quantities vs. step sizes); 

\item estimate $m_{\mathrm{FD}}$ as the slope of the regression line.

\end{enumerate*}

\item Detrended Fluctuation Analysis. DFA is one of the most used methods to determine/quantify signal's statistical self-similarity, so complexity as well. The principle is that, if the subset of an object can be rescaled to resemble statistically the original object, this object can be considered as self-similar. This implies that the self-similarity can be defined by the rescaling process \cite{Peng1995Fractal}. It is basically a modified root-mean-square analysis \cite{Ho2011prognostic}. Firstly, the time series is converted into unbounded series. Then, it is filtered by a linear-local-trend-removal to eliminate any inferences.Once this is done, root-mean-square fluctuations analysis is performed over different scales so that the scaling exponent can be characterized by the slope of the log-log plot (fluctuations vs. scales) \cite{Peng1995Quantification}. 

\item Sample Entropy (SampEn). Sample Entropy is designed to examine the regularity or fluctuations of a time series. It detects the changes in underlying episodic behavior not reflected in peak occurrences or amplitudes \cite{Pincus1992Approximateb}. If in a time series, there exist repetitive patterns of fluctuation, it will be more predictable than a time series in which such patterns are absent. So the basic idea of SampEn is to determine if similar patterns in current observation exists in the following observations. SampEn {$\ldots$ ``is precisely the negative natural logarithm of the conditional probability that a dataset of length $N$, having repeated itself within a tolerance $r$ for $m$ points, will also repeat itself for $m+1$ points, without allowing self-matches''$\ldots$} \cite{Lake2002Sample}. So, by quantifying this probability of repeatability, the regularity of the time series is examined.
\end{itemize}

The human physiological systems are very complex systems consisted of multiple organs, each of them has their own mechanical / electrophysiological properties. The interactions / interferences of these sub-systems make the output of the whole system extremely complex. So, single scale analysis of this output -- acquired signals could give global information but fail to provide more comprehensive understandings. In fact, the physiological systems exhibit nonlinear dynamics with highly irregularity or even randomness which the single scale analysis of the system output often failed to reveal the true dynamics. What's more, the cardiac arrhythmia are often associated with highly erratic fluctuations which have statistically uncorrelated noise \cite{Costa2002Multiscalea} which brings yet more challenges to single-scale analysis. Multi-scale analysis methods can overcome those shortcomings and reveal the spatial-temporal structures at multiple scales that provide more information about the system. It is thus more robust.

The scaling is performed with coarse-graining method. Its principle is to smooth the original signal, in such a way that the \emph{intrinsic dynamics} could be revealed by eliminating the local fluctuation. The lager the scale is, the smoother the obtained signal is. Given a one-dimensional discrete time series $x(t)$, ${x_1, \ldots, x_i, \ldots, x_N}$, the consecutive coarse-grained time series ${y^{(\tau)}}$, determined by the scale factor $\tau$ : $\displaystyle y^{(\tau)}_{j} = \frac{1}{\tau} \sum_{i=(j-1) \tau +1}^{j \tau} x_i, \; 1 \leq j \leq \frac{N}{\tau}$. The length of each coarse-grained time series is equal to the length of the original time series divided by the scale factor $\tau$ \cite{Costa2002Multiscalea}. The coarse-graining method is generally used in multiscale sample entropy (MSE) analysis. To compare with MSE, we performed the same scaling procedure for detrended fluctuation analysis and fractal dimension. 

\section{Results}

\subsection{Temporal Analysis : \textnormal{SDNN}}

\begin{figure}[!ht]
\subfloat[]
{\label{fig1a}	\includegraphics[width=0.49\textwidth]{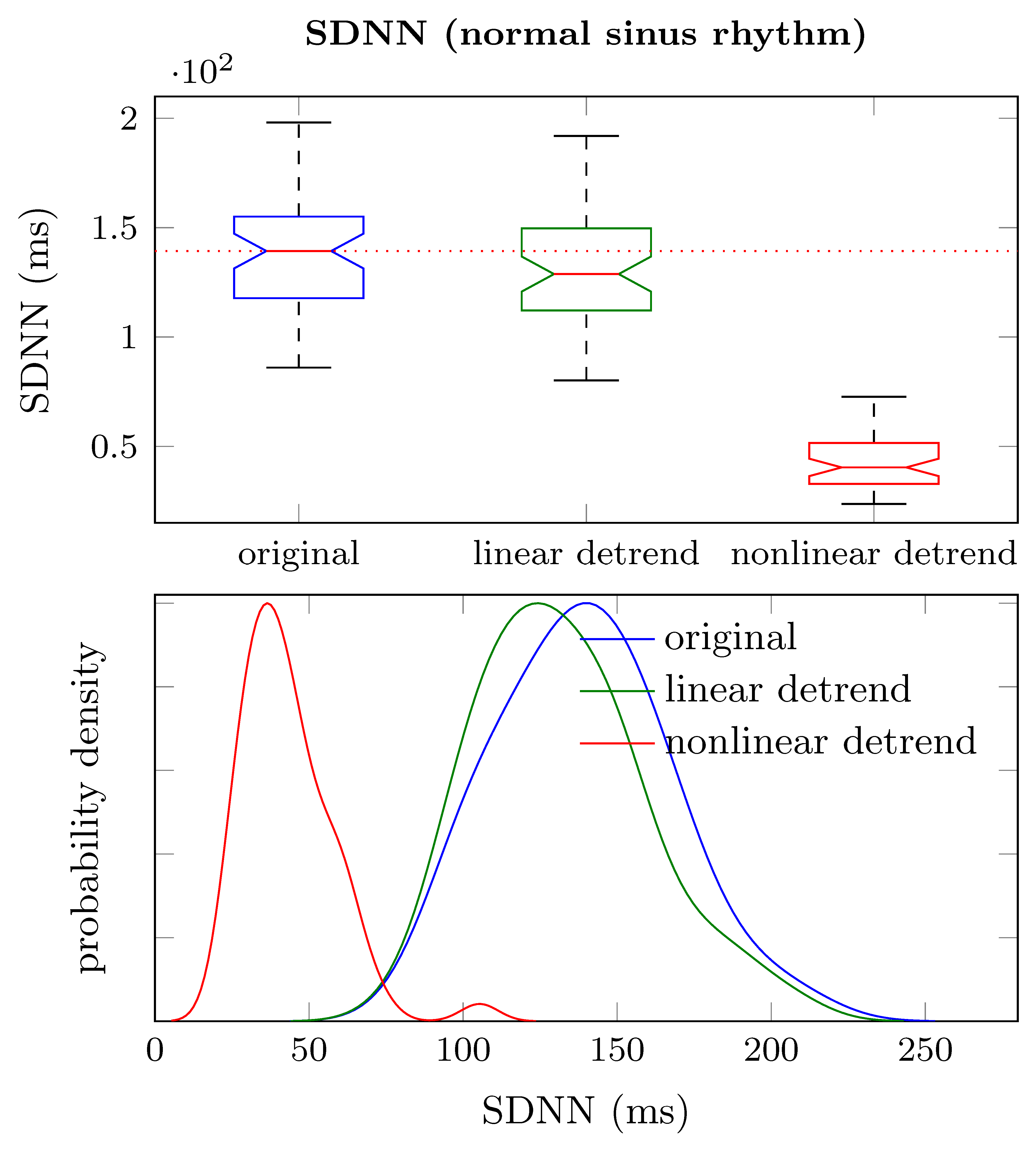}}
\subfloat[]
{\label{fig1b}	\includegraphics[width=0.49\textwidth]{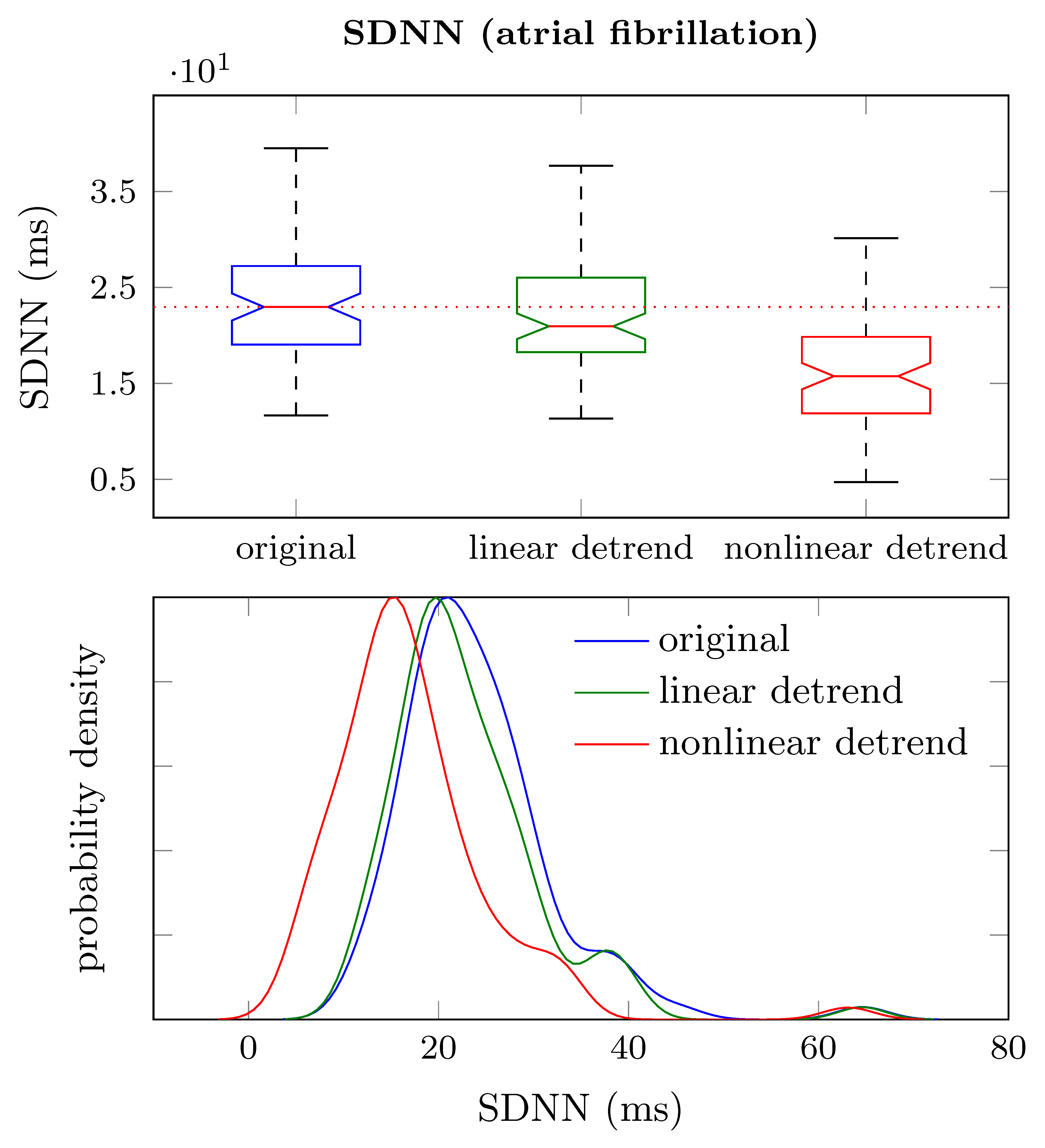}}	\\
\subfloat[]
{\label{fig1c}	\includegraphics[width=0.49\textwidth]{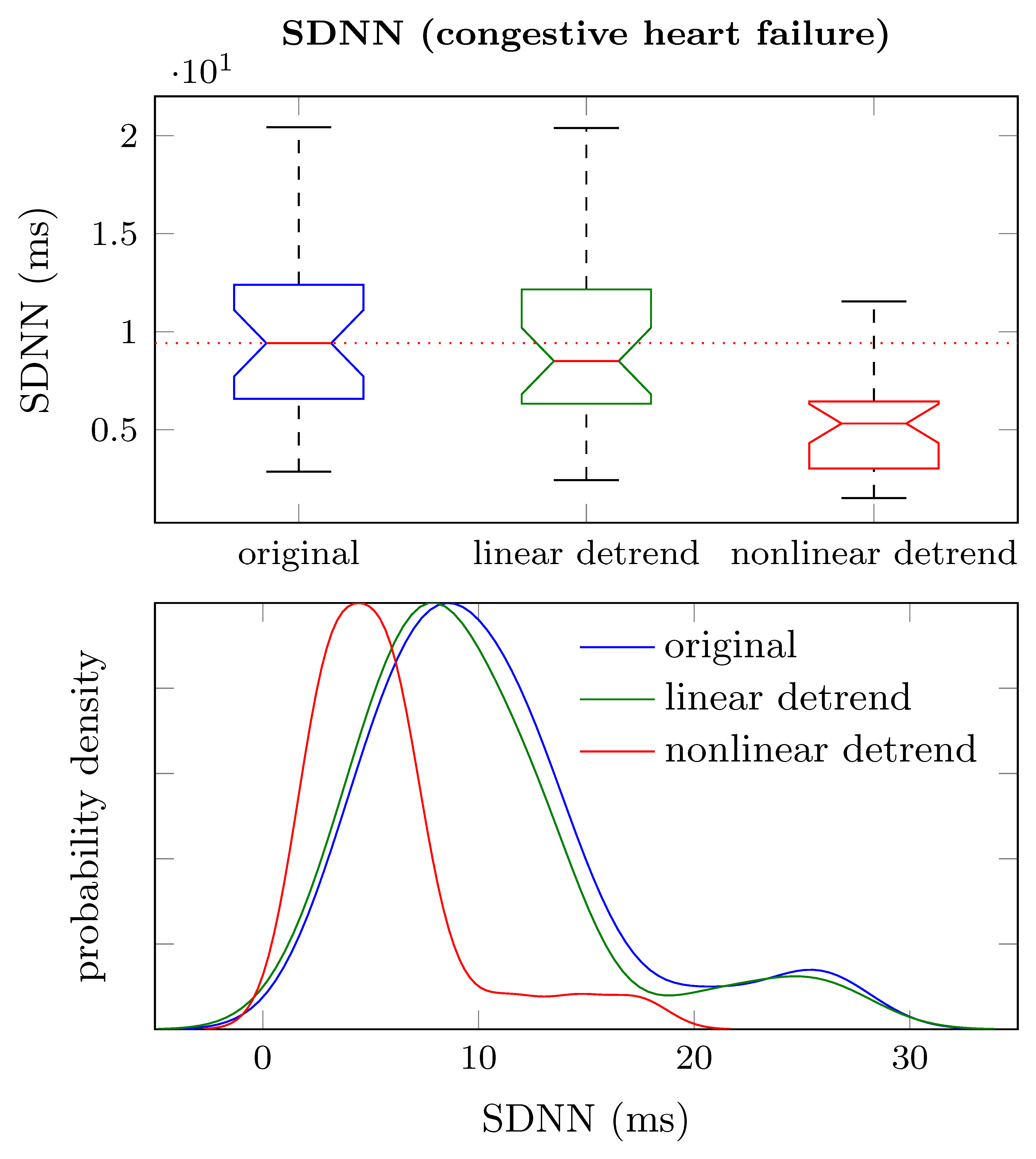}}
\subfloat[]
{\label{fig1d}	\includegraphics[width=0.49\textwidth]{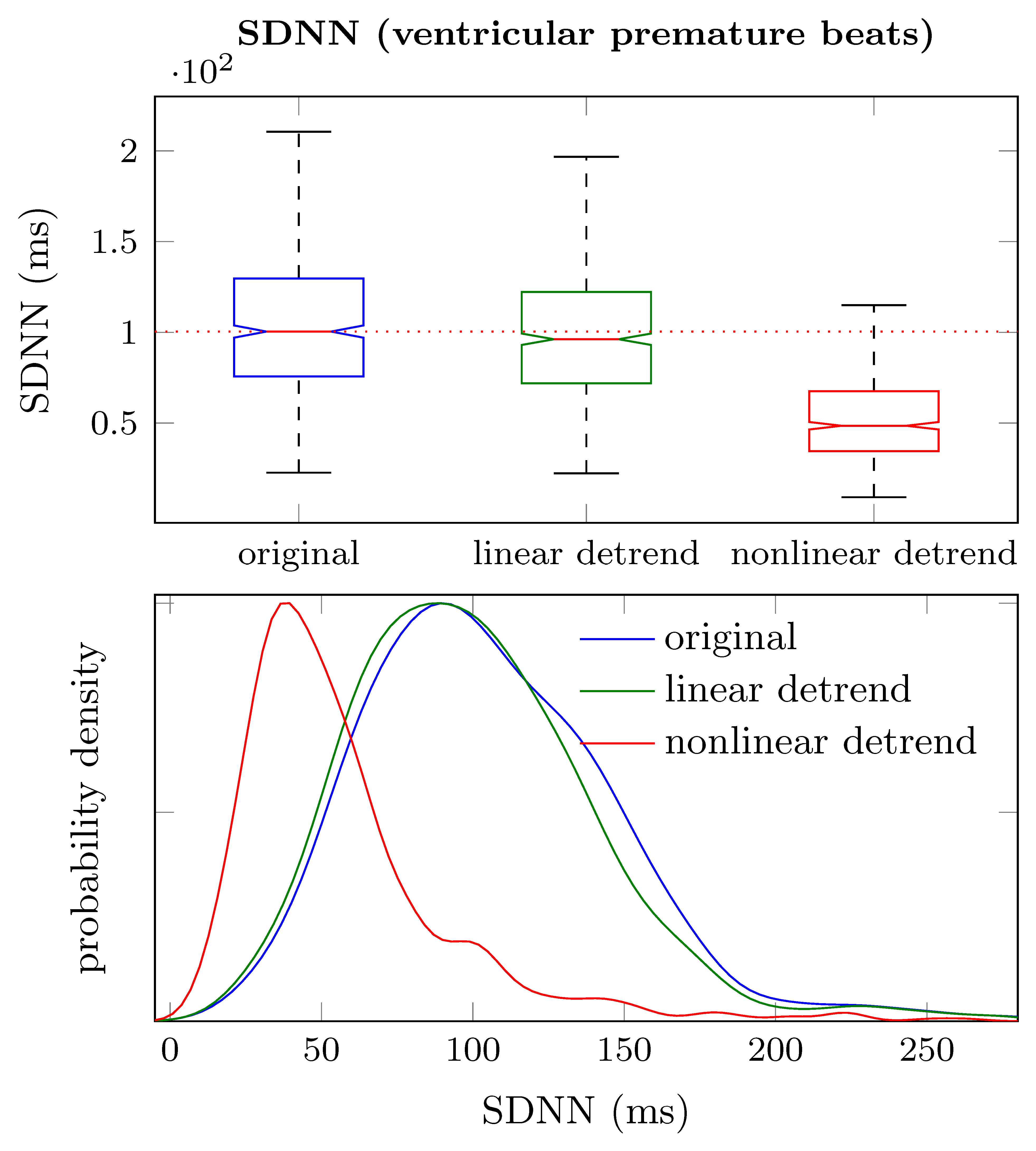}}
\caption{SDNN of HRV in original, linear trend removal and nonlinear trend removal conditions. Four types RR data : NSR, AF, CHF \& VPB.}
\label{fig_sdnn}
\end{figure}

The detrending effects on SDNN are shown in \Cref{fig_sdnn}. For better description, the results are presented in both \texttt{boxplot} and \texttt{probability density distribution}. Their probability densities are normalized for comparison reason -- these densities are from different groups, comparison of absolute densities is meaningless in this case. 

For the four RR data types, the global conclusion is the same : linear trend removal changes very little the SDNN (\Cref{fig_sdnn}). Their values are still in the same range and distribution. Comparing the SDNN values of the original RR time series and the detrended ones by Kruskal-Wallis tests, we can see that the related $p$-values are all larger than 0.01 (\Cref{table_sdnn_pv}) and showed the strong correlation. So, they can be considered from the same distribution. However, nonlinear detrending changed everything. For SDNN of NSR and VPB, their distributions are separated (\Cref{fig1a,fig1d}). The Kruskal-Wallis test on SDNN for groups of NSR and VPB showed their $p$-values are close to zero ($p<10^{-19}$, \Cref{table_sdnn_pv}), which rejected completely the correlation hypothesis of the two groups. The same observation happened for AF and CHF groups. The only difference is that the $p$-values are relatively larger, but they are still small enough ($p < 10^{-5}$, \Cref{table_sdnn_pv}) to confirm the correlation rejection. 

So, at this step, the results showed that, linear detrending does not indeed change the dynamics of the original data, only small shifting of the distribution of SDNN. The larger $p$-values showed that the SDNN for detrended data and the original data have strong correlation. Due to this correlation, the SDNN for detrended data can be regarded as an alternative to the original SDNN or improved ones in case of data classification. After the nonlinear detrending, the SDNN changed in two ways : 
\begin{enumerate*}[(i)]
\item the values are completely different from the original SDNN;
\item distribution changed as well so that this is no more correlation between the two groups.
\end{enumerate*}
It is thus certain that nonlinear detrending should not be used in SDNN analysis.

\begin{figure}[!ht]
\subfloat[]
{\label{fig2a}	\includegraphics[height=0.9\textwidth]{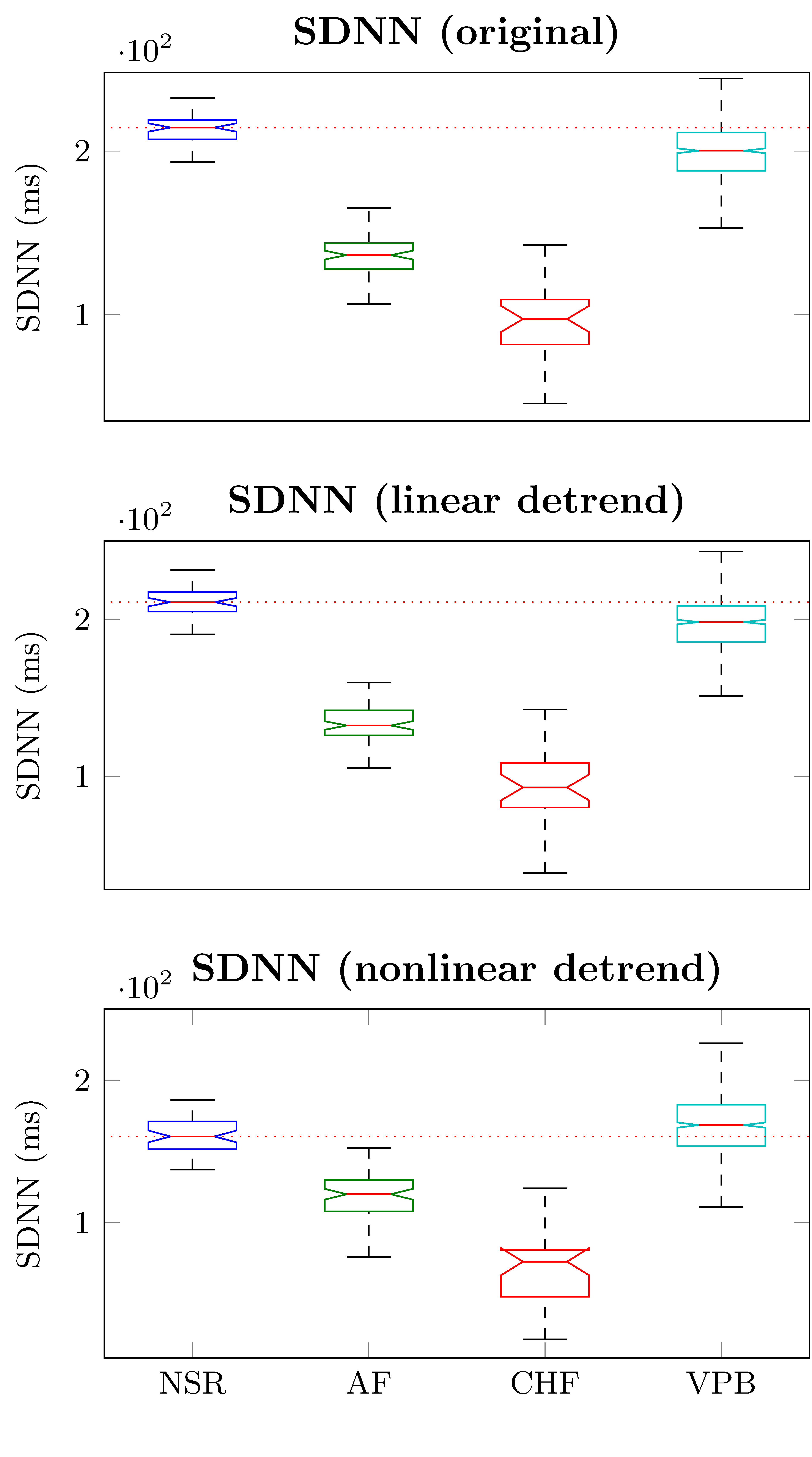}}
\subfloat[]
{\label{fig2b}	\includegraphics[height=0.9\textwidth]{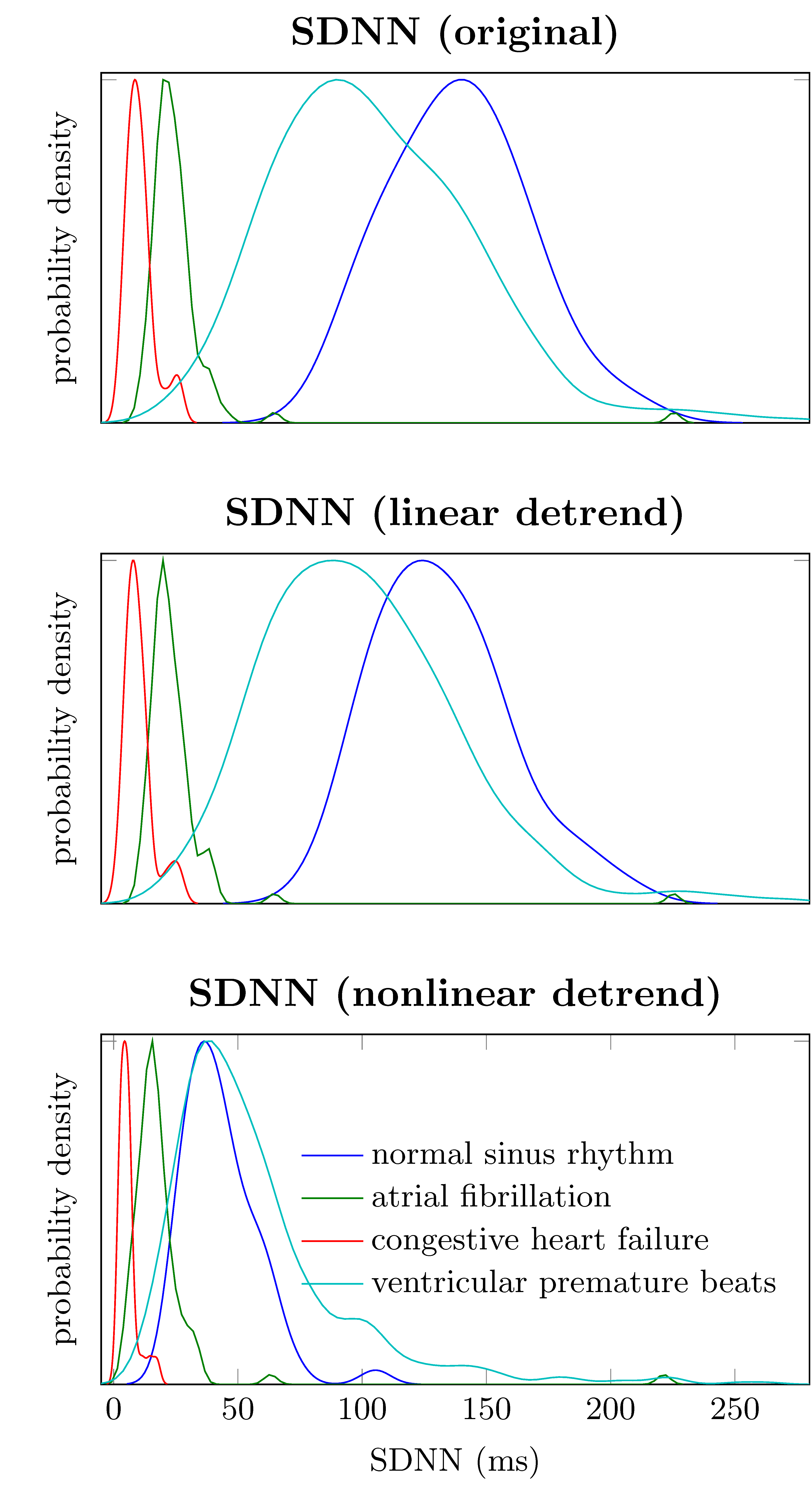}}
\caption{SDNN and their normalized probability density of four types RR data : NSR, AF, CHF \& VPB in three conditions : original data, after linear detrending and after nonlinear detrending. }
\label{fig_sdnn_hrv}
\end{figure}

We know now that the linear or nonlinear detrending effect on the SDNN. The comparisons are performed on themselves : original data vs. linearly detrended data vs. nonlinearly detrended data. How about the comparison the three pathological RR data (AF, CHF, VPB) to the baseline RR data in normal conditions ? What kind of effect by linear or nonlinear detrending on SDNN ? Comparing the SDNN of the original data (1st row, \Cref{fig_sdnn_hrv}), it is clear that SDNN of NSR is larger than any other cases (AF, CHF, VPB). This confirmed that in case of cardiac diseases, the heart losses part of its central modulation capability. So the SDNN is is then reasonably smaller. When applying linear detrending, the SDNN distributions are slightly shifted, but their order remained the same as in the first row ($\textrm{NSR} > \textrm{VPB} > \textrm{AF} > \textrm{CHF}$). Nonlinear detrending changed unfortunately this order (3rd row, \Cref{fig_sdnn_hrv}), the SDNN distribution of NSE and VPB are almost overlapped. 

\begin{table}[!ht]
\renewcommand{\arraystretch}{1.2}
\begin{center}
\begin{tabular}{r l c c }
	& & \small{original SDNN vs.} 			& \small{original SDNN vs.} \\
	& & \small{SDNN after linear detreding}	& \small{SDNN after nonlinear detreding} \\ \hline
\multirow{4}{*}{$p$-value} & \small{NSR}	&	0.1260	& 5.1279e-19	\\
						   & \small{AF}		&	0.1945	& 1.2657e-09	\\
						   & \small{CHF}	&	0.4792	& 4.1716e-05	\\
						   & \small{VDP}	&	0.0216	& 5.2836e-100	\\ \hline
\end{tabular}
\caption{$p$-values from Kruskal-Wallis tests for original SDNN vs. SDNN after linear detreding and original SDNN vs. SDNN after nonlinear detreding }
\label{table_sdnn_pv}
\end{center}
\end{table}

This section suggests that pre-processing by linear detrending RR time series has little negative effect on typical analysis in time domain. The nonlinear detrending changed a lot the analysis, especially for RR data in normal sinus rhythm and in ventricular premature beats conditions. 

\subsection{Complexity Analysis}

The SDNN in time domain gives the basic statistics / quantification of the RR time series. The analysis of nonlinear dynamics will provide yet another dimension of quantification and appropriate qualification. These nonlinear indexes are shown in \Cref{fig3,fig4,fig5,fig6,fig7,fig8}. The very first values of each curve are the basic measures, while the curves presented the multiscale analysis. 

\emph{One important note} : all the curves in the figures shown in this section are the median values of that group at each scale. They are not value for only one patient. The normal distribution tests have been performed for all values at each scale. The results showed that all these values at each scale have normal distribution. So the median value use is justified.  

\subsubsection{Multiscale Fractal Dimension analysis}

The fractal dimension $m_{\mathrm{FD}}$ of an irregular time series is always  $m_{\mathrm{FD}} \in [1,2]$. The lower limit associated with a smooth curve, and the upper limit, $m_{\mathrm{FD}} =2$, corresponding to a space-filling exceedingly rough graph. The original data and the linearly detrended data have almost the identical $m_{\mathrm{FD}}$ for the four types of RR data -- the curves are overlapped. The values of $m_{\mathrm{FD}}$ are also in the typical range as in literature. After nonlinear detrending, $m_{\mathrm{FD}}$ is changed a lot. At smaller scales, the values' change is significant for NSR, CHF and VPB data, but varied little for AF data. When increasing the scale, all is changed : the $m_{\mathrm{FD}}$ are close or equal to 2 -- typical value for random noise, this indicates that the nonlinearly detrended data changed the nature of the data. 
\begin{figure}[!ht]
\centering
\includegraphics[width=0.95\textwidth]{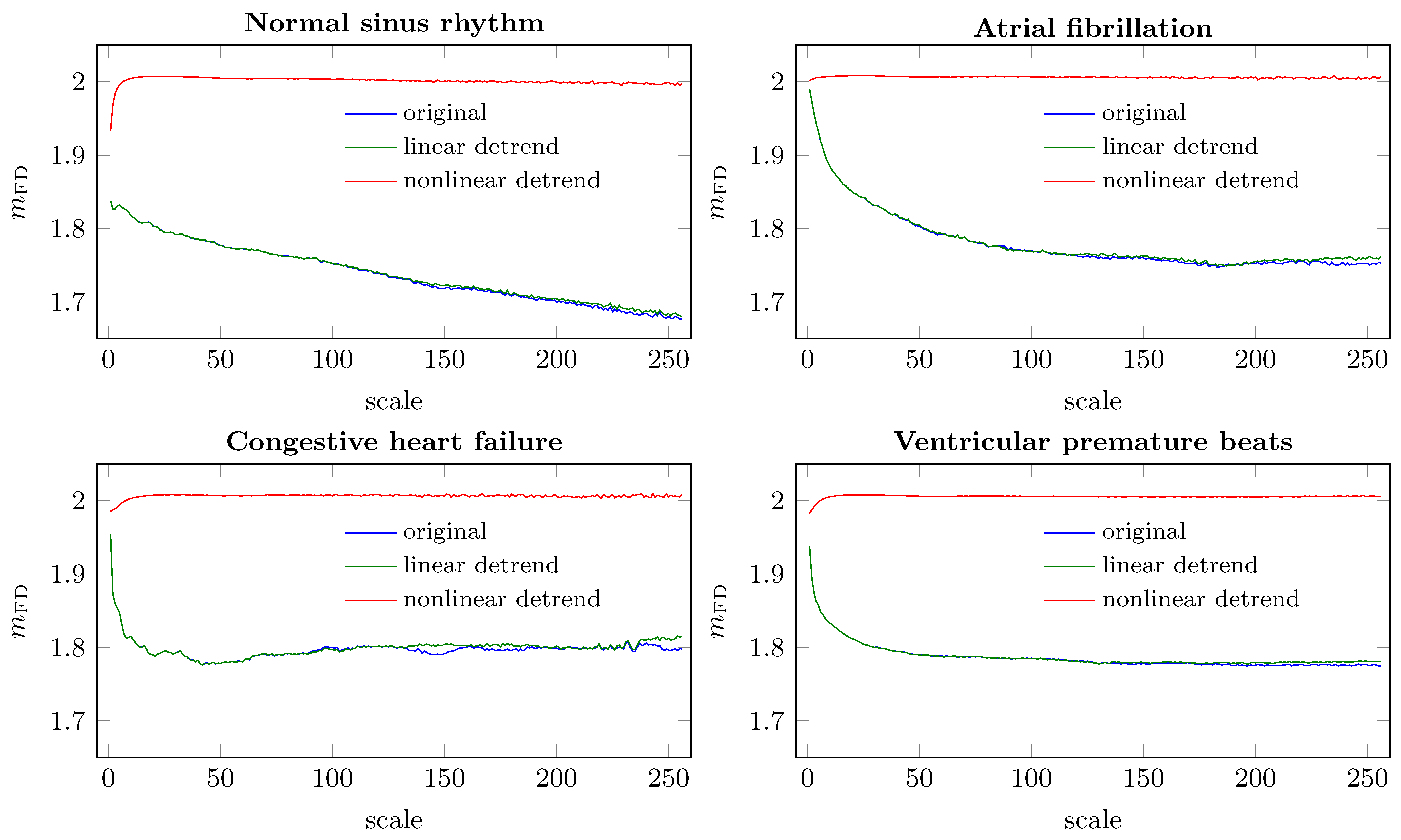}
\caption{Multiscale Fractal Dimension analysis of detrending effect on normal RR time series and pathological ones. (each curve represents median values for each group)}
\label{fig3}
\end{figure}

\begin{figure}[!ht]
\centering
\includegraphics[width=0.95\textwidth]{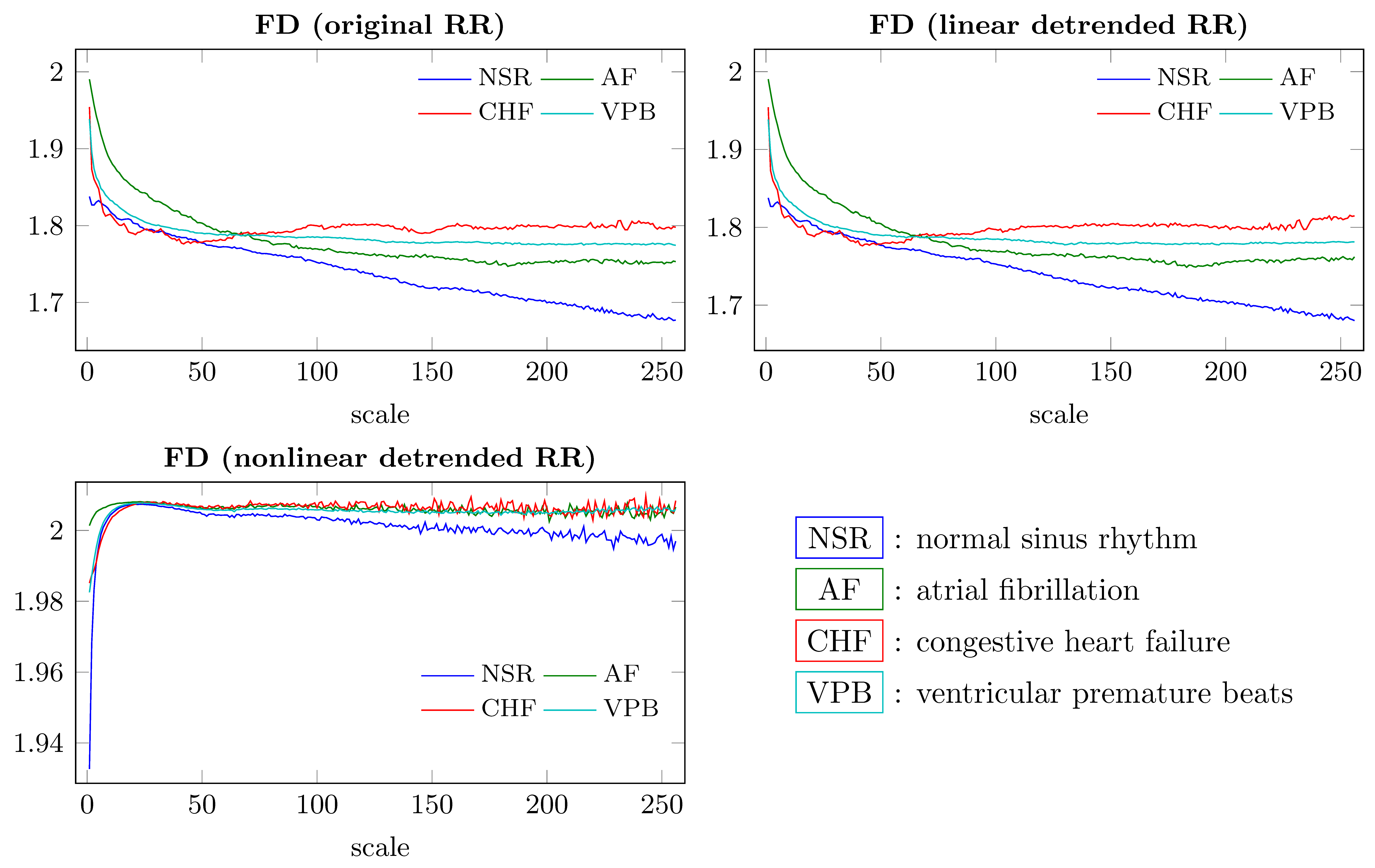}
\caption{Multiscale Fractal Dimension comparison for normal RR time series and pathological ones in three situations : normal, linear detrending and nonlinear detrending. (each curve represents median values for each group)}
\label{fig4}
\end{figure}

In \Cref{fig3}, we compared the detrending effect for each type RR data. The comparison can also be done among the $m_{\mathrm{FD}}$ for the four types of RR data in three conditions : original, linear/nonlinear detrending (\Cref{fig4}). An interesting point about the multi-scale $m_{\mathrm{FD}}$ of these data is that they have different signatures. For normal heart rate data, the $m_{\mathrm{FD}}$ is linear to scales. It is no more the case for pathological heart rate time series. The $m_{\mathrm{FD}}$ are asymptotic to certain values for each pathology. This suggested that multiscale FD could be used a marker of these diseases. 
So, these signatures are well preserved after linear detrending. Unfortunately, the nonlinear detrending made all these signatures almost indistinguishable.

\FloatBarrier
\subsubsection{Multiscale Detrended Fluctuation Analysis}

\begin{figure}[!ht]
\centering
\includegraphics[width=0.95\textwidth]{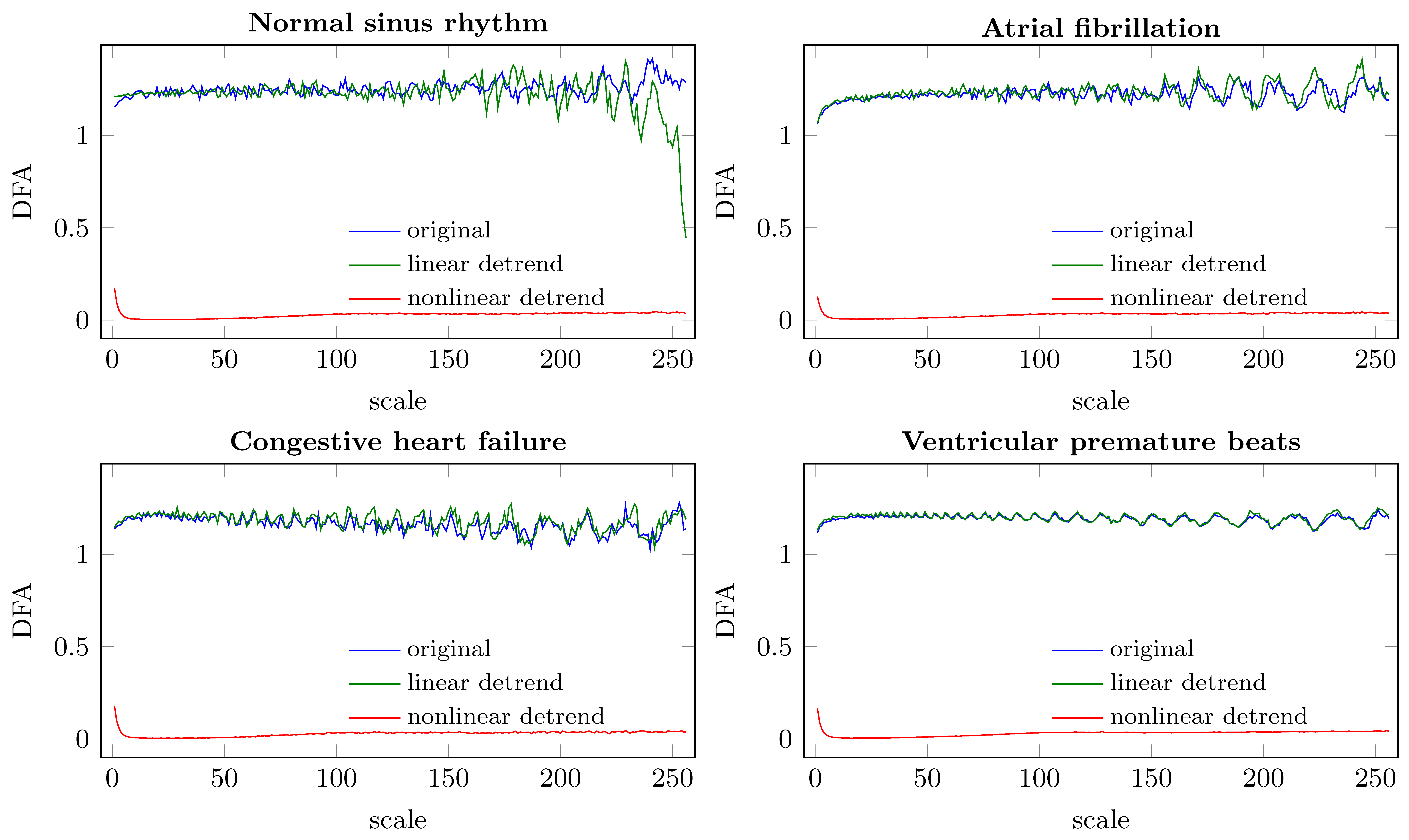}
\caption{Multiscale Detrended Fluctuation Analysis of detrending effect on normal RR time series and pathological ones. (each curve represents median values for each group)}
\label{fig5}
\end{figure}

Similar observations happened in multi-scale DFA (\Cref{fig5}). The exponent $\alpha_{\mathrm{DFA}}$ gives another quantification of the complexity. When the linearity in the data is dominant, $\alpha_{\mathrm{DFA}}$ is larger than $0.5$ (around $0.5$ for short-term correlation, $ 0.5 < \alpha \leq 1$ for persistent long-range power-law correlations). If the data is completely uncorrelated, $\alpha_{\mathrm{DFA}} = 0.5$. Outside the range of $[0.5, 1]$, $0 < \alpha < 0.5$ signals anti-persistent power-low correlations; $\alpha > 1$ means that the data is non-stationary, unbounded. The multi-scale DFA in this section confirmed the conclusion in fractal dimension analysis. The linearly detrended data has the same dynamics as the original data, the exponents $\alpha_{\mathrm{DFA}}$ are larger than 1, indicating that these data are non-stationary. However, the $\alpha_{\mathrm{DFA}}$ for nonlinearly detrended data switched completely to other side of the spectrum -- smaller than $0.5$ and even converged to zero.  This suggested strong random and anti-persistent component in the data. Since this is very likely impossible for heart rate time series, the nonlinear detrending could not be considered thus as a good choice. 

\begin{figure}[!ht]
\centering
\includegraphics[width=0.95\textwidth]{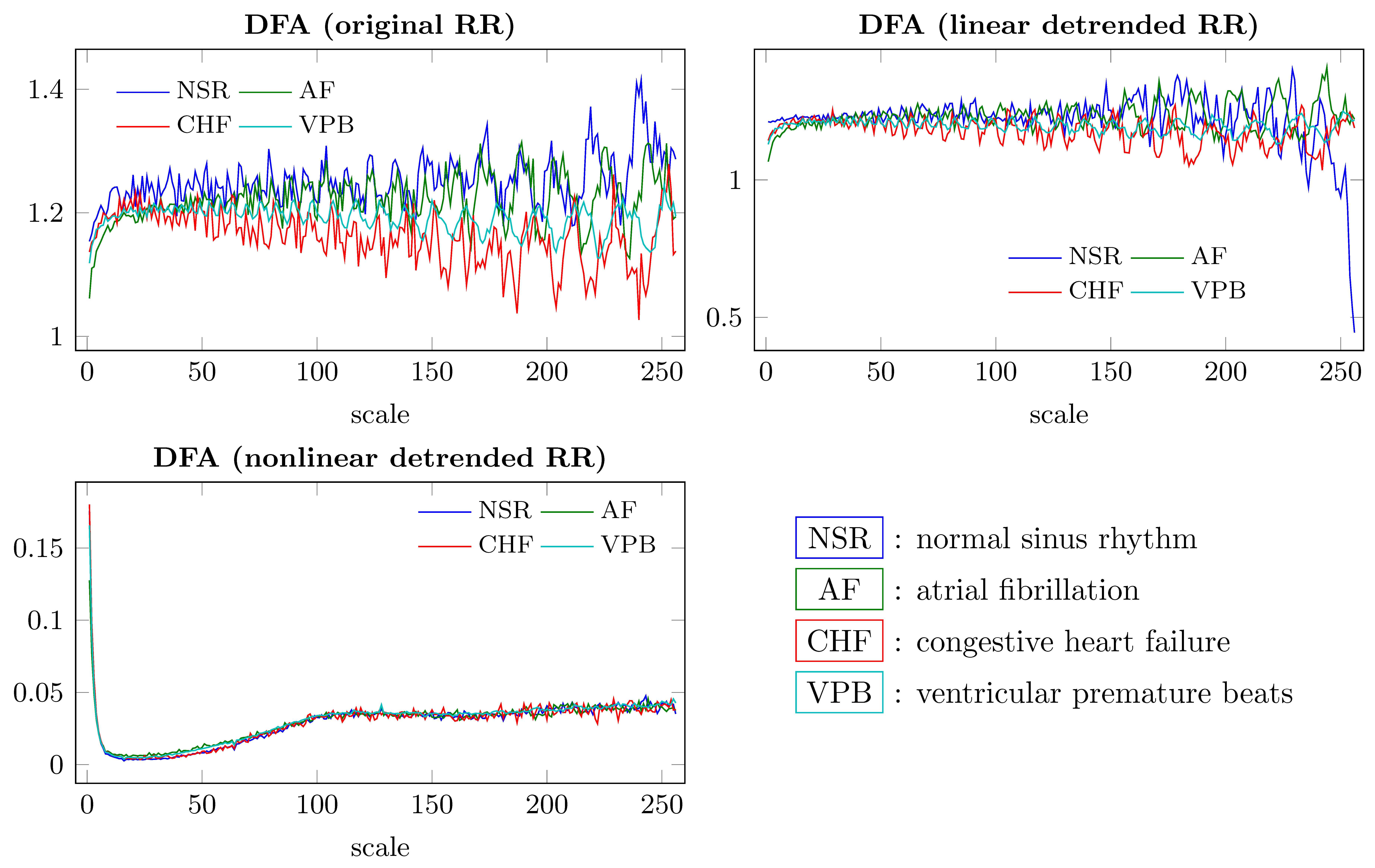}
\caption{Multiscale Detrended Fluctuation Analysis comparison for normal RR time series and pathological ones in three situations : normal, linear detrending and nonlinear detrending. (each curve represents median values for each group)}
\label{fig6}
\end{figure}

If we put the $\alpha_{\mathrm{DFA}}$ for the four RR data in the same figure, and in three conditions (normal, linnear \& nonlinear detrending), the changes are more visible (\Cref{fig6}). It showed that in original and linear detrending conditions, the difference of $\alpha_{\mathrm{DFA}}$ for each type of RR data can be still viewed. Once nonlinearly detrending the data, the $\alpha_{\mathrm{DFA}}$ for the four RR data are almost identical -- it suggested that these nonlinearly detrended data can be in fact put into the same category -- noise. It proved once again that nonlinear detrending would not be an appropriate practice. 

\FloatBarrier
\subsubsection{Multiscale Sample Entropy Analysis}

\begin{figure}[!ht]
\centering
\includegraphics[width=0.95\textwidth]{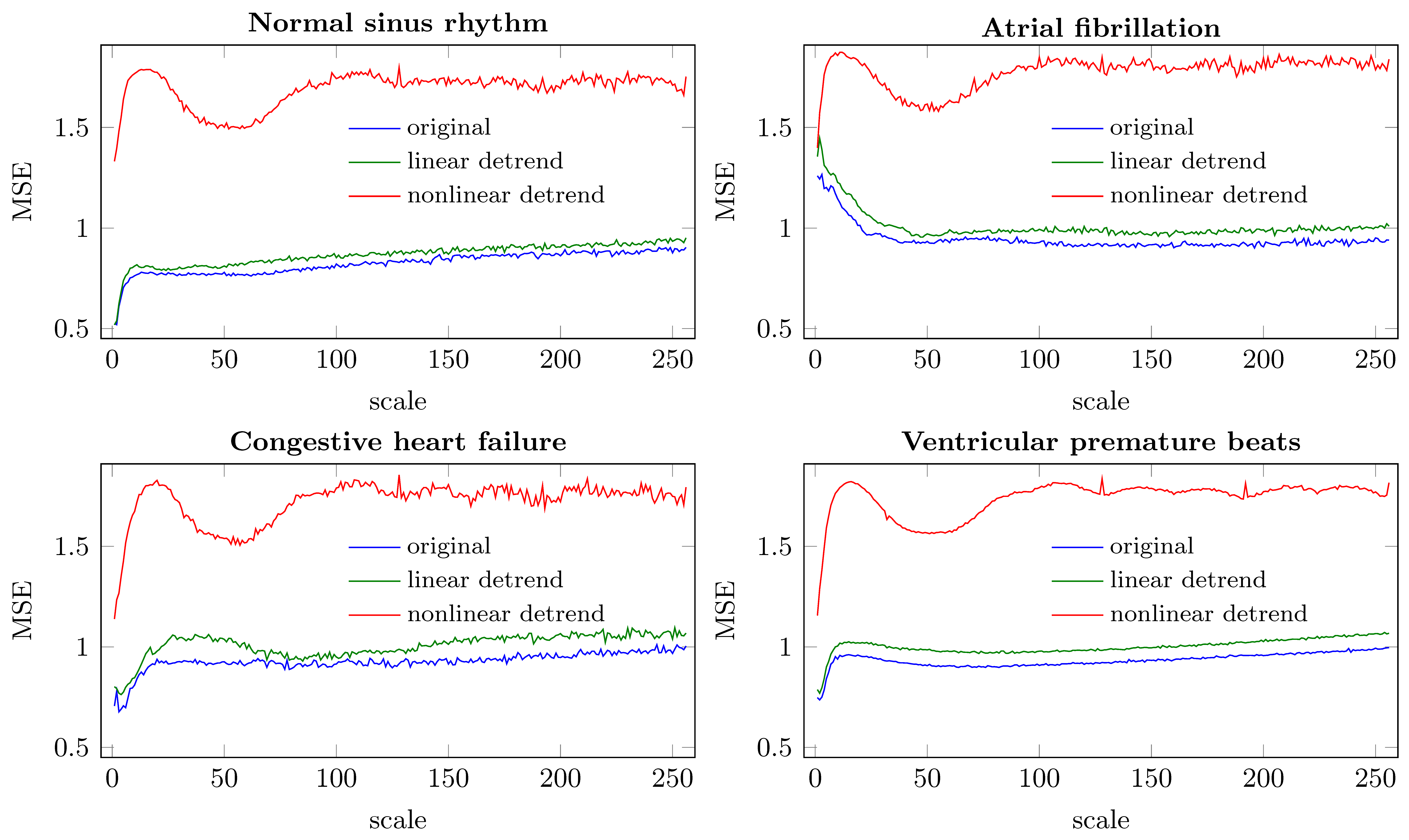}
\caption{Multiscale Sample Entropy of detrending effect on normal RR time series and pathological ones. (each curve represents median values for each group)}
\label{fig7}
\end{figure}

As above-mentioned, Sample Entropy examines the probability of similar patterns presence in the signal. If a time series contains more similar patterns, the SampEn values will be smaller; otherwise, they would be higher. It is then also a predictability term. The linear detrending changed little the MSE analysis (\Cref{fig7}), the curves (original and linear detrending) are close enough each other or just slightly shifted. Their values are smaller than those for nonlinearly detrended RR data. After nonlinear detrending, the SampEn values jumped up to 1.5 or larger and have a increasing trend. These values correspond to SampEn for random signals. 

\begin{figure}[!ht]
\centering
\includegraphics[width=0.95\textwidth]{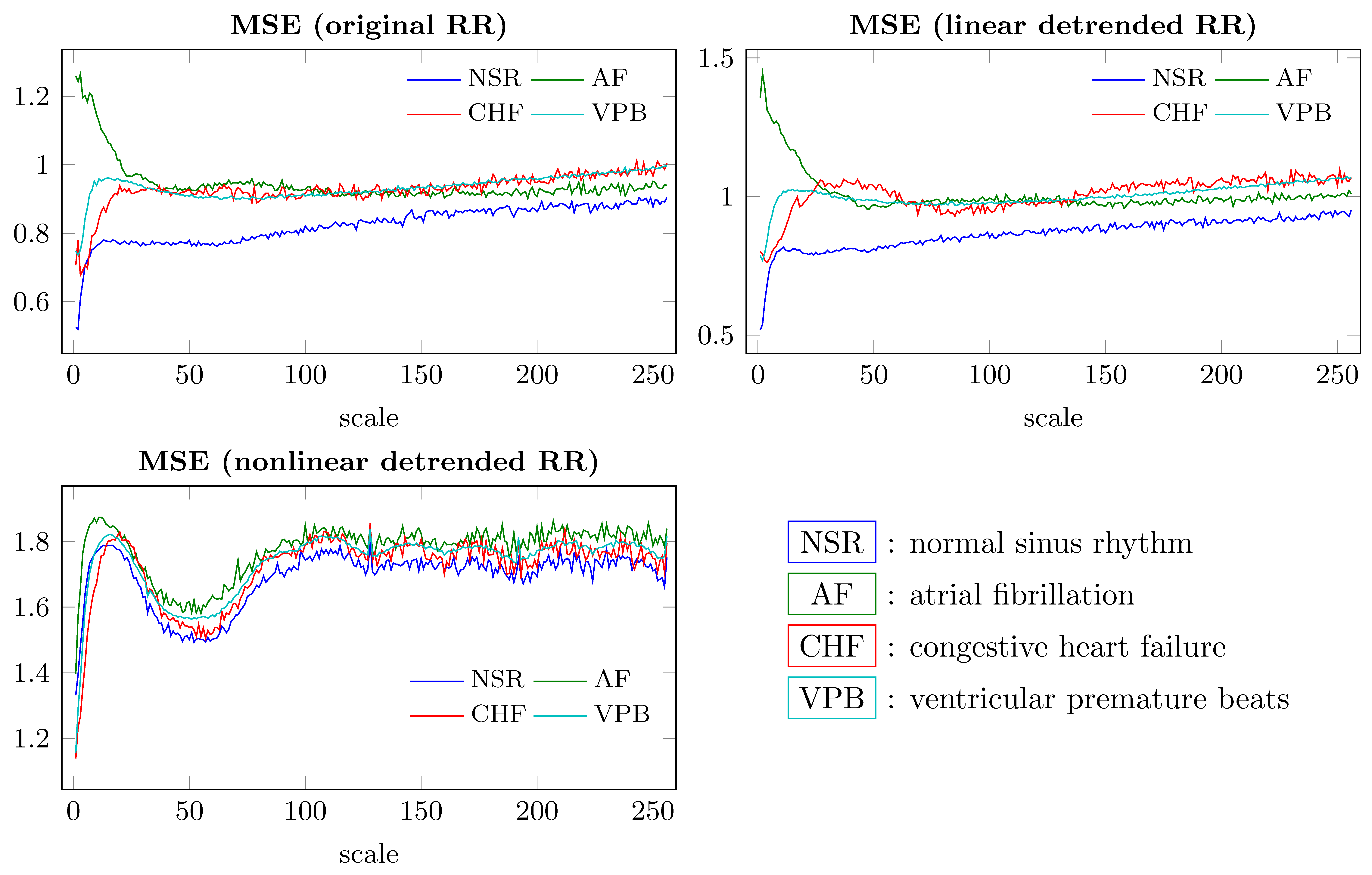}
\caption{Multiscale Sample Entropy comparison for normal RR time series and pathological ones in three situations : normal, linear detrending and nonlinear detrending. (each curve represents median values for each group)}
\label{fig8}
\end{figure}

Another observation of MSE for NSR, CHF and VPB is that the MSE for nonlinearly detrended data are completely separated from the original and linearly detrended data, including the first scales. However, in case of atrial fibrillation, the signature is different. Firstly, MSE at the very first scales are close in the three conditions (original, linear \& nonlinear detrending). Secondly, in original and linear detrending conditions, the MSE for NSR, CHF and VPB are relatively increasing and asymptotic to some thresholds. For AF, these curves are decreasing instead of increasing. 

The \Cref{fig8} showed that RR data in normal condition has smallest Sample Entropy, indicating that this type of data contains more similar patterns and is thus more predictable. In case of arrhythmia, the rhythm / harmony is broken, so the RR variation becomes less regular and less predictable, so the Sample Entropy values are larger. After the nonlinear detrending, the data are all \emph{normalized} so that the original intrinsic differences are disappeared. That's why they all looked the same, as MSE values suggested (\Cref{fig8}). 

\begin{figure}[!ht]
\centering
\includegraphics[width=0.95\textwidth]{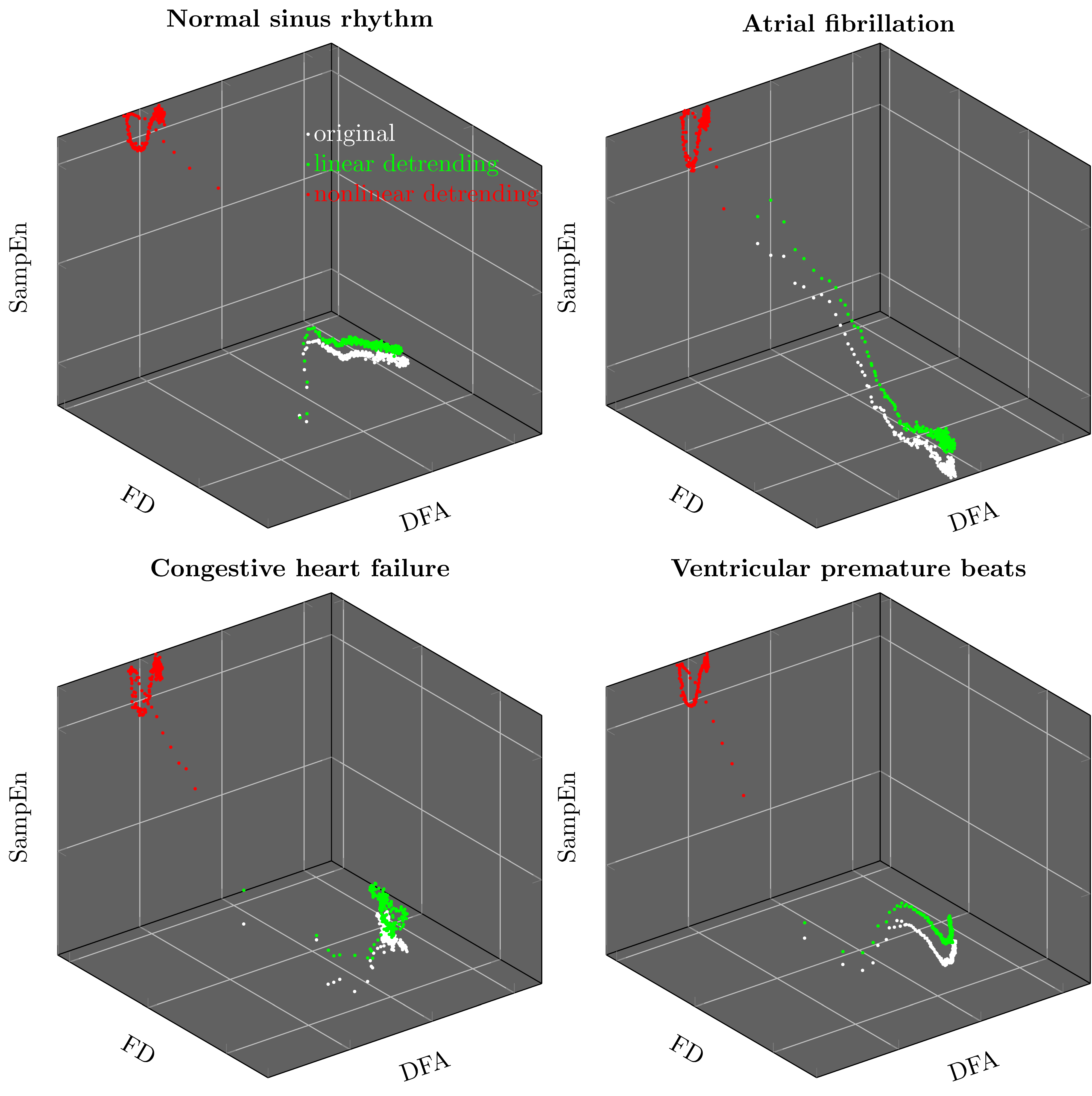}
\caption{Complexity Space based on FD, DFA \& SampEn, for RR data in NSR, AF, CHF \& VPB conditions.}
\label{fig9}
\end{figure}

Constructing a space with the three complexity indexes, as shown in \Cref{fig9}, the complexity of the nonlinearly detrended RR data situated in a position where is far from the original (and linearly detrended) data. 

Considering all previous results, in terms of either temporal analysis or nonlinear complexity analysis, it is now clear that, nonlinear detrending will change the intrinsic dynamics of the RR data. It is certain that nonlinear detrending in HRV analysis should not be advised, at least for RR data in NSR, AF, CHF \& VPB conditions or without other combined processing.

\FloatBarrier
\section{Conclusion} 

The pre-processing in biomedical signal processing played an essential but often underestimated role. The reason is that, we know very little about the biomedical system in too many cases. In fact, the acquired signals come from a \emph{black box}. The related processing is, based on very limited information, to study a high-dimensional system. In consequence, any inappropriate pre-processing would affect the fundamental dynamics. 

In case of heart rate variability analysis, the pre-processing of detrending could modify a lot the basic characteristics of the signal leading to an important change of the system. Analysis based on these modifications could be deviated. So, careful attention should be paid on this delicate procedure. 

The linear detrending affects little the global characteristics of the RR data. After linear detrending, the SDNNs are just slightly shifted and all distributions are well preserved. The cross-scale complexity (the Fractal Dimensions and exponents from Detrended Fluctuation Analysis) is almost the same as the ones for original RR data. Though there are still some differences revealed with multi-scale Sample Entropy analysis, their changes are correlated. We can conclude that pre-processing by linear detrending can be performed on RR data which does not modify the intrinsic dynamics of the data.

The same analysis with nonlinear detrending showed that this type of pre-processing could be harmful. It changed not only the SDNNs distribution, but also the order among different types of RR data. After nonlinear detrending, the SDNN for normal sinus rhythm became indistinguishable from the SDNN for RR with ventricular premature beats. This is dangerous and cannot be accepted for clinical applications. This problem can be explained by multi-scale complexity analysis. The different RR data has different complexity signature. Nonlinear detrending made the all RR data to be similar. It is thus impossible to distinguish them. In fact, the Fractal Dimension showed that nonlinearly detrended RR data has a dimension close to 2, the exponent from DFA is close to zero and SampEn is larger than 1.5 -- these complexity values are very close to those for random signal. So, when one processing completely change the nature of the data, how to draw useful / exploitable conclusion ?

This work investigated the detrending effect on the complexity analysis of heart rate. Though in HRV analysis the data used could have some influences on the conclusion, and the authors could argue that the combination with other pre-precessing techniques can avoid or bypass the detrending effect. The results are indeed not appropriate. The problem does exist. So for any clinical application, the pre-processing by detrending, if needed, should be careful conducted in order to find an optimal way which would not affect the fundamental dynamics of the original data. Unfortunately, it is hard to propose a general solution for this compensation. Because all depends on the used data. 

\FloatBarrier

\end{document}